\documentstyle[sprocl,epsfig,amsmath,amssymb,graphicx,colordvi,gnuplotcolors,feynarts,12pt]{article}

\newcommand{\gev}{\text{GeV}}
\newcommand{\tev}{\text{TeV}}

\newcommand{\MSF}{ M_{\text{Sf.}}}

\newcommand{\tb}{\tan\beta}

\newcommand{\TE}[1]{\cdot 10^{#1}}
\newcommand{\E}[1]{10^{#1}}

\newcommand{\MSSM}{{\text{MSSM}}}
\newcommand{\SM}{{\text{SM}}}
\newcommand{\LO}{{\text{LO}}}
\newcommand{\NLO}{{\text{NLO}}}

\newcommand{\NC}{{\text{NC}}}
\newcommand{\CC}{{\text{CC}}}
\newcommand{\XC}{{\text{XC}}}

\def\be{\begin{equation}}
\def\ee{\end{equation}}
\def\bea{\begin{eqnarray}}
\def\eea{\end{eqnarray}}

\begin{document}

\title{Radiative Corrections to
Deep-Inelastic 
Neutrino--Nucleon 
Scattering
in the MSSM
}

\author{O. BREIN$^1$ \footnote{Speaker.}, B. KOCH$^{2,3}$ and W.
HOLLIK$^4$}

\address{$^1$ Institut f\"ur Theoretische Physik E, RWTH Aachen,
    D-52056 Aachen, Germany\\
$^2$ Institut f\"ur Theoretische Physik,
Johann Wolfgang Goethe-Universit\"at,\\
 D-60054 Frankfurt am Main, Germany\\
$^3$ Frankfurt International Graduate School for Science (FIGSS)\\
 D-60054 Frankfurt am Main, Germany\\
$^4$ Max-Planck-Institut f\"ur Physik,
     F\"ohringer Ring 6, D-80805 M\"unchen, Germany}


\maketitle\abstracts{
We discuss the radiative corrections to charged and neutral current
deep-inelastic neutrino--nucleon scattering in the minimal supersymmetric
standard model (MSSM).
In particular, deviations from the Standard Model prediction for the ratios
of
neutral- to charged-current cross sections, $R^\nu$ and $R^{\bar\nu}$, are
studied, and results of a scan over the 
MSSM parameter space are presented.
}

\section{Introduction}
\begin{figure}[ht]
\begin{center}
{
\unitlength=0.3mm%
\begin{picture}(300,60)(0,20)
\begin{footnotesize}

\begin{feynartspicture}(300,100)(3,1)
\FALabel(44.,91.96)[]{}
\FADiagram{}
\FAProp(0.,15.)(10.,14.)(0.,){/Straight}{1}
\FALabel(4.84577,13.4377)[t]{$\nu_\mu$}
\FAProp(0.,5.)(10.,6.)(0.,){/Straight}{1}
\FALabel(5.15423,4.43769)[t]{{$\begin{array}{c}
d,\;\bar{u}\\
s,\;\bar{c}
\end{array}$}}
\FAProp(20.,15.)(10.,14.)(0.,){/Straight}{-1}
\FALabel(14.8458,15.5623)[b]{$\mu$}
\FAProp(20.,5.)(10.,6.)(0.,){/Straight}{-1}
\FALabel(15.1542,7.56231)[b]{{$\begin{array}{c}
u,\;\bar{d}\\
c,\;\bar{s}
\end{array}$}}
\FAProp(10.,14.)(10.,6.)(0.,){/Sine}{-1}
\FALabel(8.93,10.)[r]{$W$}
\FAVert(10.,14.){0}
\FAVert(10.,6.){0}

\FADiagram{}

\FADiagram{}
\FAProp(0.,15.)(10.,14.)(0.,){/Straight}{1}
\FALabel(4.84577,13.4377)[t]{$\nu_\mu$}
\FAProp(0.,5.)(10.,6.)(0.,){/Straight}{1}
\FALabel(2.15423,4.43769)[t]{$u,\;d,\;s,\;c$}
\FAProp(20.,15.)(10.,14.)(0.,){/Straight}{-1}
\FALabel(14.8458,15.5623)[b]{$\nu_\mu$}
\FAProp(20.,5.)(10.,6.)(0.,){/Straight}{-1}
\FALabel(17.7542,7.)[b]{$u,\;d,\;s,\;c$}
\FAProp(10.,14.)(10.,6.)(0.,){/Sine}{0}
\FALabel(8.93,10.)[r]{$Z$}
\FAVert(10.,14.){0}
\FAVert(10.,6.){0}
\end{feynartspicture}
\end{footnotesize}
\end{picture}
}
\end{center}
\caption{\label{born}
Tree-level Feynman graphs for neutral-current and charged-current
scattering of muon-neutrino and quark.}
\end{figure}
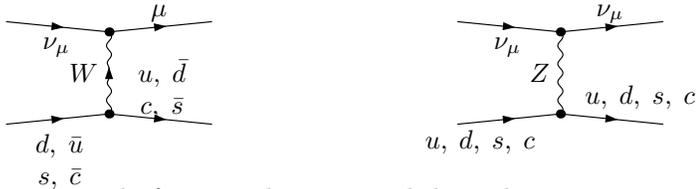

In the Standard Model (SM), neutral (NC) and charged current (CC) neutrino--nucleon
scattering are described in leading order by $t$-channel $W$ and $Z$ exchange, respectively
(see Fig.~\ref{born}).
At the NuTeV experiment, $\nu_\mu$ and $\bar\nu_\mu$ beams of a mean energy
of 125 GeV were scattered
off a target detector and the ratios
$R^\nu = {\sigma_\NC^\nu}/{\sigma_\CC^\nu}$ and
$R^{\bar\nu} = {\sigma_\NC^{\bar\nu}}/{\sigma_\CC^{\bar\nu}}$
were measured.
The NuTeV collaboration also provided a determination of the on-shell weak
mixing angle~\cite{nutev02},
\begin{align*}
\sin^2\theta_w^{\text{\tiny on-shell}} & = 0.2277 \pm 0.0013 (stat.)
\pm 0.0009 (syst.)\;.
\end{align*}
This value is about $3\sigma$ below the value derived from the residual set
of precision observables~\cite{sinthetaSM}.
The analysis 
makes use of the
Paschos-Wolfenstein relation~\cite{PW},
\begin{align*}
R^- & = \frac{R^\nu-rR^{\bar\nu}}{1-r} = \frac{1}{2} - \sin^2\theta_w
+\text{corrections}\;, &
r & = \frac{\sigma_\CC^{\bar\nu}}{\sigma_\CC^\nu} \approx \frac{1}{2} \;,
\end{align*}
and the measurement of counting rates
$R^\nu_{\text{exp}}, R^{\bar\nu}_{\text{exp}}$.
From the theoretical ratios $R^{\nu}, R^{\bar{\nu}}$ for a
fixed beam energy, one obtains
the SM prediction for the experiment 
$R^\nu_{exp}(SM), R^{\bar{\nu}}_{exp}(SM)$
by a detailed Monte Carlo (MC) simulation.
These predictions differ from the counting rates
$R^\nu_{exp}, R^{\bar{\nu}}_{exp}$ by
{\cite{nutev-DeltaRnu}} 
\begin{align*}
\Delta R^\nu & = R^\nu_{\text{exp}} - R^\nu_{\text{exp}}(SM)
= -0.0032 \pm 0.0013 \;,\\
\Delta R^{\bar\nu} & = R^{\bar\nu}_{\text{exp}} -
R^{\bar\nu}_{\text{exp}}(SM)
= -0.0016 \pm 0.0028 \;.
\end{align*}
%
\smallskip

Basically, there are three types of explanations of the observed anomaly.\\
(a) {\em There is no signal.} The result might be a statistical fluctuation
or theoretical errors may have been underestimated.
This line of thought lead to re-analyses of the electroweak 
radiative corrections to neutrino--nucleon DIS \cite{DDH}, 
which have been originally calculated almost 20 years ago
\cite{bardin-dokuchaeva}.\\
(b) {\em The apparent signal is due to neglected but relevant SM effects.}
An asymmetry between the distribution of the strange quark 
and its corresponding anti-quark in the nucleon ($s \not= \bar s$)
could account for part of the deviation.
The result of the measurement of the weak mixing angle could
also be influenced by 
a violation of the usually assumed isospin symmetry ($u_p \not= d_n$),
or nuclear effects which have not been taken into account. 
For a list of other sources see refs.~\cite{davidson,sm-explanations} and
references therein.\\
(c) {\em The signal is due to new physics.}
Many suggestions have been made, like effects from 
modified gauge boson interactions (e.g. in extra dimensions), 
non-renormalizable operators,
leptoquarks (e.g. $R$-parity violating SUSY) 
and SUSY loop effects (e.g. in the MSSM), just to name a few of them (see
\cite{davidson,shufang} and
references therein for an overview).

Here we consider the effects of 
the radiative corrections to 
neutrino--nucleon DIS in the MSSM.
Although the NuTeV anomaly is not  settled as yet,
it is interesting to check how far the MSSM could account for such a
deviation.
There are two earlier studies in the 
literature~\cite{davidson,shufang}. 
Both treat the loop effects in the limit of zero-momentum transfer
of the neutrino to the hadron, neglecting kinematical cuts and 
potential effects from the parton distribution functions.
They conclude
that the radiative corrections in the MSSM cannot be made
responsible for the NuTeV anomaly, owing to the wrong sign.

Our calculation includes various kinematical effects. In particular,
we include the full $q^2$ dependence of the one-loop amplitudes,
evaluate hadronic cross sections using PDFs
\cite{Martin:1995ws}
cuts on the hadronic energy in the final state 
($20\,\gev < E_{\text{had.}} < 180\,\gev$)
at the mean neutrino beam energy of 125 GeV.
Moreover, we perform  
a thorough parameter scan for the 
radiative corrections $\delta R^{\nu(\bar\nu)}$
over the relevant MSSM parameter space.

\section{MSSM radiative corrections to $\nu_\mu N$ DIS}

%
The difference between the MSSM and SM predictions,
$\delta R^n = R^n_\MSSM - R^n_\SM$ with
$R^n = {\sigma^n_\NC}/{\sigma^n_\CC}$,
with
$
(\sigma^n_\XC)_\NLO  = (\sigma^n_\XC)_\LO + \delta\sigma^n_\XC 
\; (\text{X = N,C} ; n = \nu,\bar\nu)
$,
can be expanded as follows,
\begin{align*}
\delta R^n & = \left(\frac{\sigma^n_\NC}{\sigma^n_\CC}\right)_\LO 
 \left(\frac{(\delta\sigma^n_\NC)_\MSSM-(\delta\sigma^n_\NC)_\SM}{(\sigma^n_\NC)_\LO}
-\frac{(\delta\sigma^n_\CC)_\MSSM-(\delta\sigma^n_\CC)_\SM}{(\sigma^n_\CC)_\LO}
\right) .
\end{align*}
Thus, only differences between MSSM and SM radiative corrections and 
leading-order (LO) cross-section expressions appear.
$R$-parity conservation in the MSSM makes
the Born cross section the same as in the SM (up to a negligible 
extra contribution involving a virtual charged Higgs boson).
Consequently, 
contributions from real photon emission
and all SM-like radiative corrections without virtual Higgs bosons
are equal in the MSSM and the SM.
Therefore, the difference between the MSSM and SM prediction  for the 
quantity $R^n$ 
at one-loop  order in the electroweak corrections boils down
to the genuine superpartner (SP) loops and the difference between 
the Higgs-sector contributions, i.e.\ schematically given by
\begin{align}
\label{delR-schema}
\delta R^n & \propto 
\big(\; [\text{SP loops}]
+ [\text{Higgs graphs MSSM}-\text{Higgs graphs SM}] \;\big) \;.
\end{align}
%
%
The second term in eq.~(\ref{delR-schema}) vanishes
when the MSSM Higgs sector is SM-like, which is the case for 
a $CP$-odd Higgs mass $m_A \gtrsim 250\,\gev$.
The superpartner-loop contributions for CC (anti-)neutrino--quark scattering
consist of $W$-boson selfenergy insertions, loop contributions to 
the $Wq\bar q'$- and $W\nu_\mu \mu^- (W\bar\nu_\mu \mu^+)$-vertex, and
box-graphs with double gauge-boson exchange.
Analogous contributions appear in the NC case with $W$ replaced by
$Z$ and $\mu^-(\mu^+)$ by $\nu_\mu(\bar\nu_\mu)$. 
Additionally, there are SP-loop contributions to the 
photon--$Z$ mixing selfenergy.
The partonic processes were calculated using the 
computer programs FeynArts and FormCalc \cite{FAFC}.

\section{MSSM parameter scan for $\delta R^{\nu(\bar\nu)}$}
\begin{figure}[tb]
{\setlength{\unitlength}{1cm}
\begin{picture}(15,8)(0,0)

\put(-1.5,3){\resizebox*{.43\width}{.43\height}{\includegraphics*{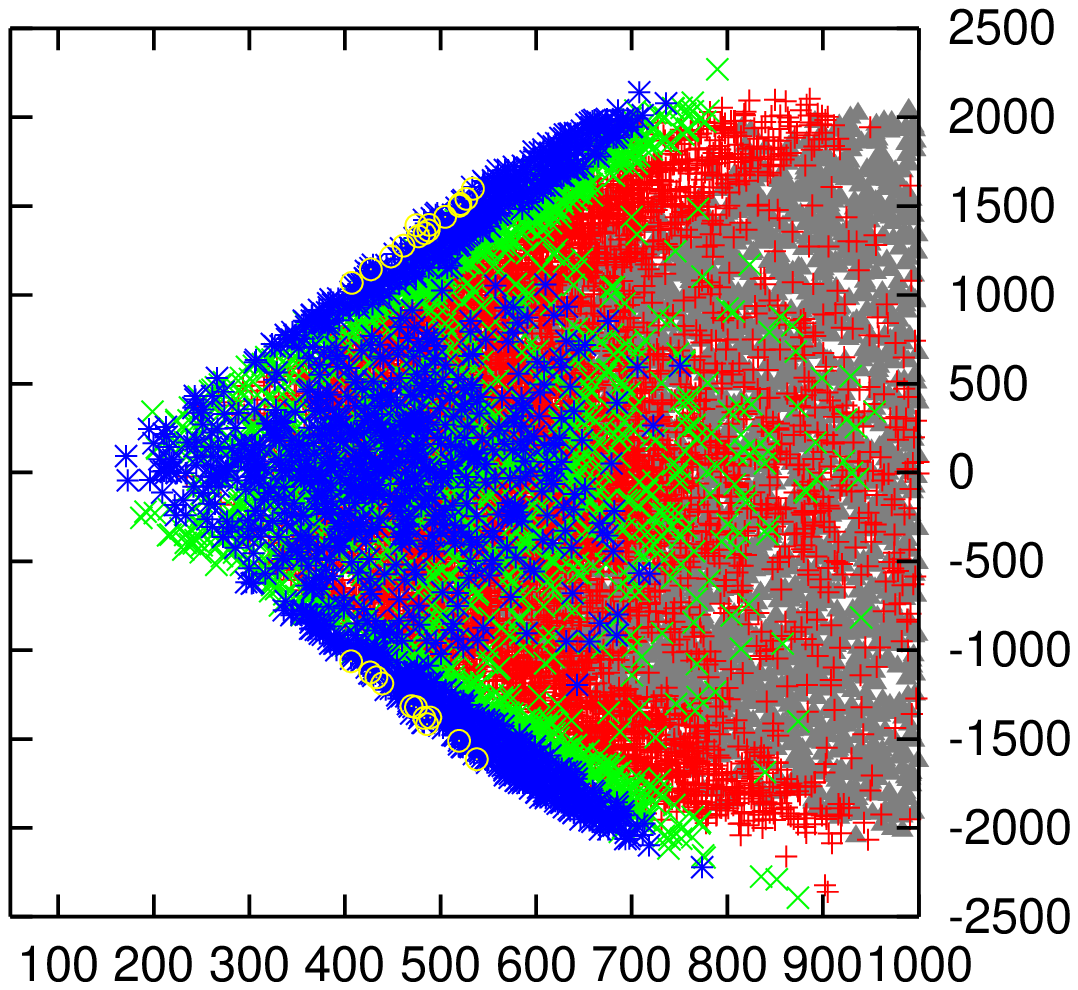}}}
\put(.6,8.4){(a)}
\put(.6,7.9){$\delta R^\nu$}
\put(-1.5,-1.4){\resizebox*{.43\width}{.43\height}{\includegraphics*{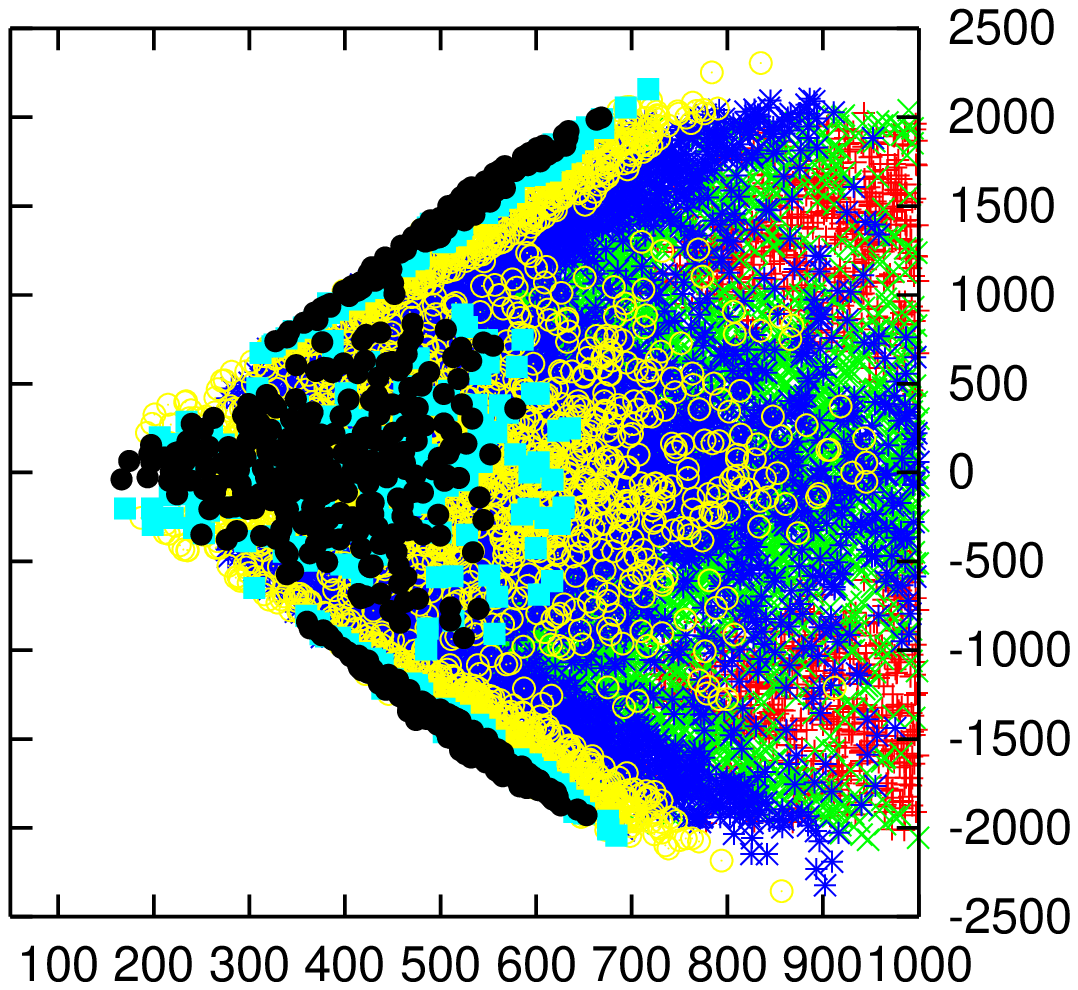}}}
\put(.6,4){(b)}
\put(.6,3.5){$\delta R^{\bar\nu}$}
\put(5.2,0.4){\resizebox*{.43\width}{.43\height}{\includegraphics*{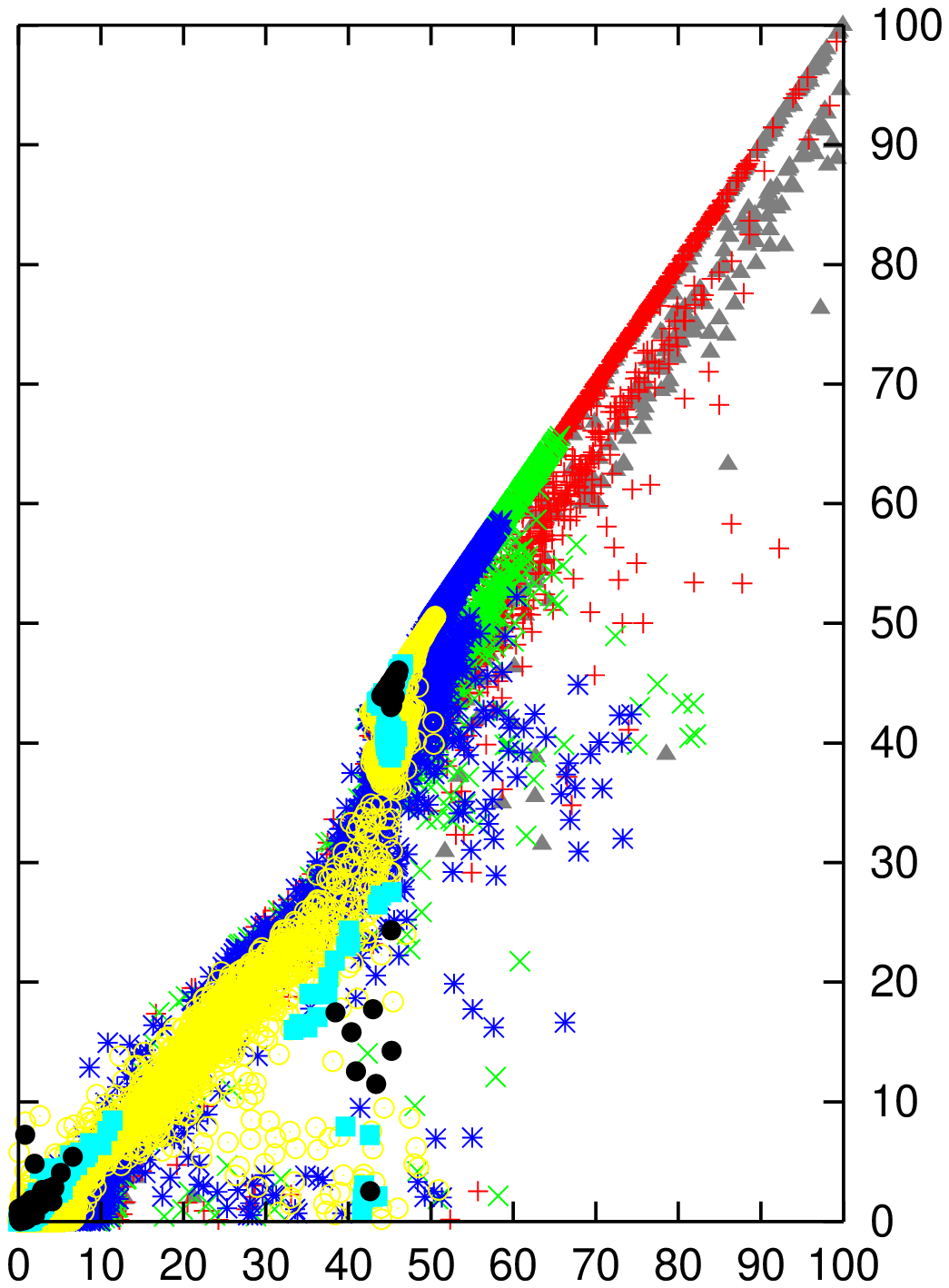}}}
\put(7.5,8.4){(c)}
\put(7.5,7.9){$-\delta R^{\nu}$}

\put(7.5,2.5){\tiny $\delta R^{\nu(\bar\nu)}$}
\put(7,2.2){\tiny\GNUPlotI{$0 - 1\TE{-5}$}}
\put(7,1.9){\tiny\GNUPlotA{$1\TE{-5} - 5\TE{-5}$}}
\put(7,1.6){\tiny\GNUPlotB{$5\TE{-5} - 1\TE{-4}$}}
\put(7,1.3){\tiny\GNUPlotC{$1\TE{-4} - 3\TE{-4}$}}
\put(7,1.0){\tiny\GNUPlotF{$3\TE{-4} - 8\TE{-4}$}}
\put(7,0.7){\tiny\GNUPlotE{$8\TE{-4} - 1\TE{-3}$}}
\put(7,0.4){\tiny\GNUPlotG{$> 1\TE{-3}$}}

\put(2,-.2){\small $\MSF\;[\gev]$}
\put(5.,6.8){\small $X_t\;[\gev]$}
\put(5.,2.4){\small $X_t\;[\gev]$}
\put(11.5,6){\small $m_{\chi^0_1}\;[\gev]$}
\put(9.2,2.5){\small $m_{\chi^\pm_1}\;[\gev]$}
\end{picture}
}
\caption{\label{drnu}
Results of the parameter scan (i) (see text) for (a) $\delta R^\nu$ and
(b) $\delta R^{\bar\nu}$ in the $\MSF$--$X_t$--plane. Panel (c)  
shows $-\delta R^\nu$ of all results with $\delta R^\nu < 0$ 
of the scan (ii) in the $m_{\chi^\pm_1}$--$m_{\chi^0_1}$--plane. 
The color code for the displayed values applies
to all three panels.
}
\end{figure}

We make use of a technique described in \cite{adaptive-scan}
which allows us to perform parameter scans with emphasis on 
specific features of the prediction.
%
We perform two different parameter scans over the following set of MSSM
parameters 
in the ranges given:
$50\,\gev  \leq M_1, M_2, M_3, \MSF \leq 1\,\tev$, 
$1 \leq \tb \leq 50$, $-2\,\tev \leq \mu, A_t, A_b \leq 2\,\tev$,
where $M_1, M_2, M_3$ are gaugino mass parameters,
$\MSF$ is a common sfermion mass scale,
$\tb$ is the ratio of the two Higgs vacuum expectation values in the
MSSM,
$\mu$ is the supersymmetric Higgs mass term,
$A_t$ and $A_b$ are soft-breaking trilinear couplings.
We choose $m_A = 500\,\gev$, such that the Higgs sector is SM-like
and the 
superpartner loops determine $\delta R^{\nu(\bar\nu)}$.
%
In a first step~(i),  we scan the quantities $\delta R^{\nu(\bar\nu)}$ 
with an emphasis on large deviations 
from the SM and neglect all points in parameter space
which violate 
mass exclusion limits for Higgs bosons
or for superpartners or the $\Delta\rho$-constraint on sfermion mixing.
Fig.~\ref{drnu}(a) and \ref{drnu}(b) show the values  $\delta R^{\nu}$ and 
 $\delta R^{\bar\nu}$, respectively, 
resulting from our parameter scan
in the $\MSF$--$X_t$--plane ($X_t = A_t - \mu/\tb$).
There are 
large radiative corrections for a kind of 
''large-mixing'' scenario ($|X_t| \approx 3 \MSF$).
Interestingly, we obtain only positive values for 
$\delta R^{\nu}$  and $\delta R^{\bar\nu}$.
Therefore, in a second scan (ii) we neglect all constraints and 
put an emphasis on negative values of  $\delta R^{\nu(\bar\nu)}$.
This scan adaptively improves on sampling parameter points where
$\delta R^{\nu(\bar\nu)}$ is negative,
as it is suggested by the NuTeV result,
and finds indeed such points.
But, all of them
violate at least one of the above-mentioned constraints
and are thus excluded.
Fig.~\ref{drnu}(c) shows only parameter points with
$\delta R^{\nu} < 0$
resulting from the second scan
in the plane of the lightest neutralino and chargino mass
($m_{\chi^0_1}, m_{\chi^\pm_1}$), 
which proves to be decisive for the sign of 
$\delta R^{\nu(\bar\nu)}$. 
Not shown in Fig.~\ref{drnu}(c) is a narrow area of 
negative values with $|\delta R^\nu| < \E{-5}$ 
around  $m_{\chi^0_1} \approx m_{\chi^\pm_1}$
extending up to $\approx 500\,\gev$. 
Fig.~\ref{drnu}(c) shows that negative values with a magnitude $>
\E{-4}$
only appear for $m_{\chi^0_1}, m_{\chi^\pm_1} < 80\,\gev$.
From this analysis we conclude that the NuTeV result for $R^\nu$ and
$R^{\bar\nu}$ cannot be explained by MSSM 
radiative corrections in parameter regions 
where the superpartner loops dominate.

\section{Summary}
The NuTeV measurement of $\sin^2\theta_w$ is 
intriguing but has to be further established.
Especially, confirmation by other experiments is desirable.
Loop effects from the non-standard particles in the MSSM
do not provide aviable explanation of the deviation
observed by NuTeV in the electroweak $\sin^2\theta_w$.
The size of the deviation could be of the right order, but
it either appears with the wrong sign or violates other electroweak
constraints.
For a final detailed  analysis, also the differences between the
Higgs sectors of the SM and the MSSM have to be incorporated.
In any case, 
interesting restrictions on the
MSSM parameters can be obtained.


\end{document}